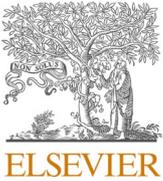
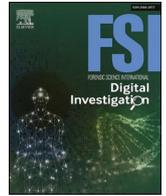

DFRWS EU 2024 - Selected Papers from the 11th Annual Digital Forensics Research Conference Europe

# DFRWS EU 10-year review and future directions in Digital Forensic Research

Frank Breitinger [a,*], Jan-Niclas Hilgert [b], Christopher Hargreaves [c], John Sheppard [d], Rebekah Overdorf [a], Mark Scanlon [e]

[a] School of Criminal Justice, Faculty of Law, Criminal Justice and Public Administration, University of Lausanne, 1015 Lausanne, Switzerland
[b] Fraunhofer FKIE, Bonn, Germany
[c] Department of Computer Science, University of Oxford, United Kingdom
[d] Department of Computing and Mathematics, South East Technological University, Waterford, Ireland
[e] Forensics and Security Research Group, School of Computer Science, University College Dublin, Ireland

## ARTICLE INFO

## ABSTRACT

Conducting a systematic literature review and comprehensive analysis, this paper surveys all 135 peer-reviewed articles published at the Digital Forensics Research Conference Europe (DFRWS EU) spanning the decade since its inaugural running (2014–2023). This comprehensive study of DFRWS EU articles encompasses sub-disciplines such as digital forensic science, device forensics, techniques and fundamentals, artefact forensics, multimedia forensics, memory forensics, and network forensics. Quantitative analysis of the articles' co-authorships, geographical spread and citation metrics are outlined. The analysis presented offers insights into the evolution of digital forensic research efforts over these ten years and informs some identified future research directions.

## 1. Introduction

The first Digital Forensic Research Workshop (DFRWS) was held in the USA in 2001 and produced the seminal report: "A Road Map for Digital Forensic Research" (Palmer et al., 2001), which has informed digital forensics research for over two decades. Since then, the not-for-profit DFRWS organisation was formed and now organises three digital forensic academic conferences each year; DFRWS USA, EU (est. 2014), and APAC (est. 2021).

The DFRWS EU conference has significantly contributed to the digital forensic research community in Europe and globally through many publications, keynotes, workshops, tutorials, panels, and community discussions and networking. However, the priorities of the DFRWS EU research community have not been studied. Consequently, this paper serves as a review of the accepted full research papers published in the ten years of proceedings of DFRWS EU 2014–2023 to analyse the impact of the conference's output. As part of the discussion, potential future research directions and best practices are discussed, aiming to increase the discipline's research efforts and foster its development into a more mature scientific domain.

*Related work*. There are other trend analysis papers for digital forensics, with "Digital forensics research: The next 10 years" by Garfinkel (2010) being the most cited. Since this initial work from 2010, a similar study has been published by Luciano et al. (2018). Other works took a slightly different approach and considered trends, such as number of publications, primary vs. secondary research, citation counts, country of origin, academia vs industry, type of publication, and keywords used (Dezfoli et al., 2013; Baggili et al., 2013; Caviglione et al., 2017; Horsman and Aney Biju Mammen, 2020). Given all the review articles, Casino et al. (2022) decided to do a review of reviews where they reviewed 109 review papers and 51 reports and identified seven research areas in digital forensics. They also included a year-wise analysis of the literature reviewed. The most recent article is by Reedy (2023) and includes references to 260 papers from multiple journals. While DFRWS EU was not cited specifically, it did include those from the special DFRWS proceedings issues of FSI: Digital Investigation. Focusing a survey on a specific conference provides highly contextualised insights and encourages community engagement, while tracking the conference's evolution and its impact on the field. In addition, this approach facilitates tailored recommendations and insights. With DFRWS EU's decade milestone, a longitudinal study spanning this ten-year period for this particular venue emerges as a significant point of interest within the community.

* Corresponding author.
E-mail addresses: frank.breitinger@unil.ch (F. Breitinger), hilgert@cs.uni-bonn.de (J.-N. Hilgert), christopher.hargreaves@cs.ox.ac.uk (C. Hargreaves), john.sheppard@setu.ie (J. Sheppard), rebekah.overdorf@unil.ch (R. Overdorf), mark.scanlon@ucd.ie (M. Scanlon).

*Contributions.* This paper makes the following contributions.

- A systematic review and categorisation of all 135 full research DFRWS EU articles from 2014 to 2023.
- Quantitative analysis and visualisation of the articles' metadata including co-authorship patterns, geographical spread, and citation metrics.
- Potential future directions are suggested to aid digital forensic researchers in setting their research agendas.

*Article Outline.* First, the methodology used is outlined, followed by the Quantitative Insights and the Qualitative Findings. Based on the analysis, some Future Directions: Research and Best Practices are outlined. The last two sections provide the Limitations and the Conclusion.

## 2. Quantitative and qualitative methodology

The relatively small dataset size necessitated a focus on content for qualitative and metadata for quantitative analysis, maximizing the value of each approach.

*Article Collection.* Accepted articles are published open access in a special issue of Elsevier's Forensic Science International: Digital Investigation (FSI: DI, formerly Digital Investigation). However, authors may opt-out of Elsevier publication. Consequently, the majority of articles were accessed via the Elsevier website and the non-Elsevier articles were added manually. Extended abstracts, posters, etc., were not included.

### 2.1. Quantitative analysis methodology

Elsevier's Article and Abstract APIs were used to collect quantitative information on all available papers including author names, affiliations, institutions, countries, and article keywords, abstracts, and text. This dataset was then augmented with the total and yearly citation data, which is accessible via Google Scholar. The data were collected on Sept. 5, 2023.

*Dataset Limitations.* One paper (Gruber et al., 2023a) was not listed on Google Scholar. Four DFRWS self-published, i.e., non-Elsevier, papers (Gruber et al., 2023a; Schneider et al., 2021; Ottmann et al., 2022; Coates and Breitinger, 2022) were added manually to the analysis.

### 2.2. Qualitative analysis methodology

**Step 1 - Screening:** Articles were sorted by year for initial screening by an individual. This step aimed to create, gather, and assign preliminary tags, e.g., theme and content. Each article was given a main category, potential tags, and identified trends. For example, Boztas et al. (2015) produced:

DFRWS session title: Investigating New Hardware
Main category: Device forensics
Other possible tags: Small-scale device forensics, artefact, data acquisition, data analysis
Possible trends: Experiment

**Step 2 - Classification creation:** The gathered main categories and possible tags were generalised to narrow down this exhaustive list. This resulted in a classification, which is explained in the next step.
**Step 3 - Article organisation:** Each article was reviewed and summarised. Based on the summary and classification, articles were then re-tagged to match one of the categories, and then re-reviewed with this categorical context.
**Step 4 - Analysis and synthesis:** Lastly, the collected information was analysed and identified common themes, patterns, or trends across the literature.

*Classification.* The need for a new classification arose from the unsuitability of existing ones. Existing, focused taxonomies, such as data fragment classification techniques by Poisel et al. (2014, Fig. 1), and memory acquisition techniques by Latzo et al. (2019), are complemented by broader ones like IoT Forensics by Yaqoob et al. (2019, Fig. 4). However, these are still insufficient. Comprehensive taxonomies like by Casino et al. (2022) and Wu et al. (2020) cover domains such as multimedia, cloud, IoT, blockchain, and computer, but neglect aspects like digital forensic process or evidence validation. Consequently, the classification developed, shown in Fig. 1, differs from existing ones. It does not claim to encompass all of digital forensic science, but reflects a decade of DFRWS EU publications.

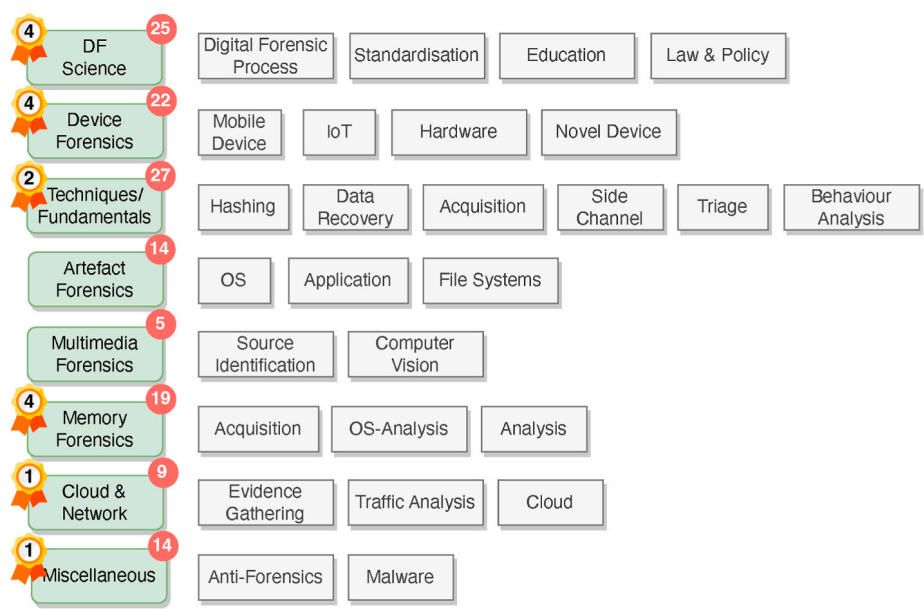

**Fig. 1.** Categorisation of DFRWS EU published work with indications of the numbers of papers and best paper awards per category.

# 3. Quantitative insights

## 3.1. Citations

Fig. 2 shows yearly citations for each DFRWS EU paper as of Sept. 5, 2023. The graph is dominated by the two most cited papers from Karbab et al. (2018) and Servida and Casey (2019). Excluding 2023 papers, all received at least one citation. The average and median citations were 23.6 and 13.0, respectively. 73.3 % of papers had 5 or more citations, 57.0 % had 10 or more, 11.9 % had 50 or more, and 2.2 % had 100 or more.

## 3.2. Author countries

Fig. 3 illustrates the diversity of DFRWS EU author affiliations by country. While most authors are based in European institutions, particularly Germany (aside from 2015), there are consistent contributions from outside Europe. In 2023, non-European authors even outnumbered European. The graph also highlights the COVID-19 pandemic's effect on geographical diversity (2020–2022 saw fewer non-European papers), and the impact of DFRWS APAC's establishment in 2020 (noted by a decrease in Asian-based papers).

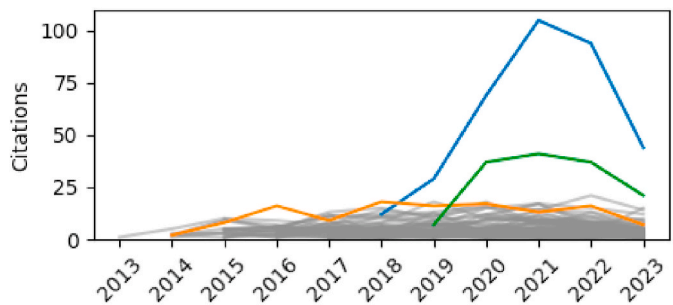

**Fig. 2.** Citations per year. Each line represents a single paper and the number of citations it receives per year. The coloured lines are the three papers with the highest citations.

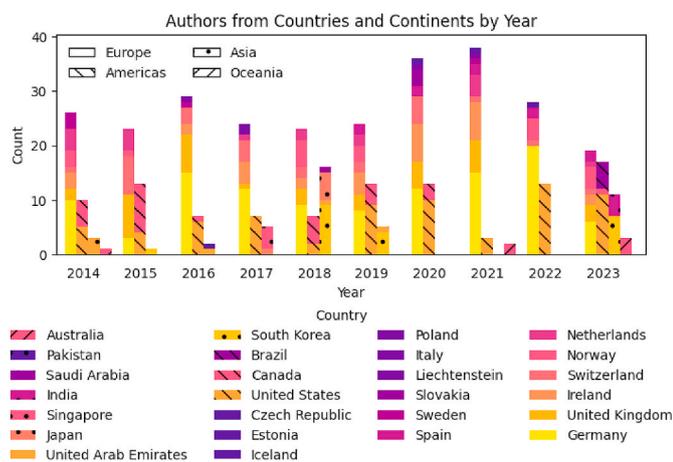

**Fig. 3.** The number of authors per affiliation country per year. The left-most bar for each year represents the number of authors from Europe, the next from the Americas, etc. Note that this graph does not take into account the relatively rare case of multiple papers by the same author in a single year. Such authors are intentionally counted multiple times in the graph, since the aim is to understand where authors come from *by paper* – not on an individual level.

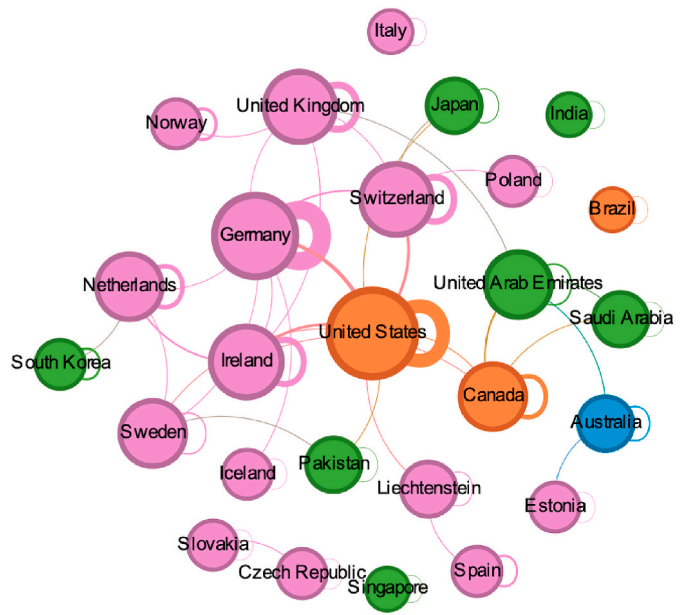

**Fig. 4.** A graph (Bastian et al., 2009) demonstrating the co-authorship according to the country of the authors' institutions. All authors of the paper (not only the primary) were considered and for authors with multiple affiliations in different countries, the first listed was chosen. Each node represents a country $\mathscr{C}_i$ and an edge between nodes $\mathscr{N}_{\mathscr{C}_i}$ and $\mathscr{N}_{\mathscr{C}_j}$ indicates that an author from $\mathscr{C}_i$ co-authored a paper with an author from $\mathscr{C}_j$. The edges are weighted by the number of collaborations. Note that all countries have self-loops, indicating that authors from the same countries collaborated on the same paper. Nodes are coloured by continent. Node size is determined by degree.

Fig. 4 indicates significant collaboration between authors from different countries and continents, despite most submissions coming from Europe. While many cross-institution papers originate within the same country, particularly Germany and the US, the graph's diverse non-self connections reveal numerous cross-country collaborations.

## 3.3. Author collaborations

Fig. 5 displays DFRWS EU author collaborations. Most clusters are small and isolated, indicating single papers by authors with no other DFRWS EU papers. Some of the larger clusters are interconnected — Green (centred around node A) to Blue (containing nodes G, E, and H) and Gold (centred around node C) to Purple (containing nodes B, F, and D). The largest two clusters have 31 nodes. The first is centred around a single author, Felix Freiling (A). The gold cluster shares this structure, centred around Mourad Debbabi (C). Some smaller clusters follow a related "bow tie" pattern, with one main author having 2 papers with 2 sets of co-authors. The other largest cluster, in purple, has several larger nodes (B - Mark Scanlon, D - Nhien-An Le-Khac, and F - Elias Bou-Harb) linking many smaller ones, indicating regular collaboration within the cluster. The blue cluster has the same structure, although with fewer connections. The three larger nodes are G - Harald Baier, E – Frank Breitinger, and H - Ricardo J. Rodríguez. The large pink cluster at the top with no central node stems from a single paper (Park et al., 2018) with authors not involved in other DFRWS EU papers. The majority of the remaining clusters result from single-paper publications.

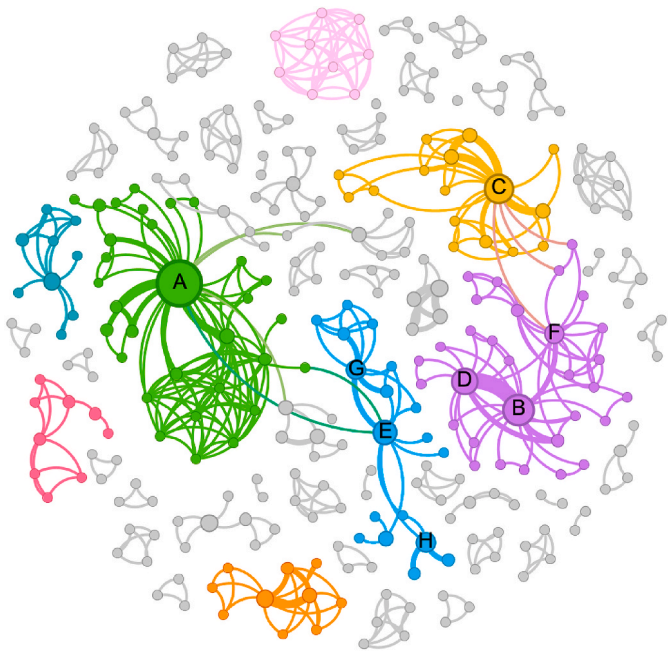

**Fig. 5.** A graph (Bastian et al., 2009) exhibiting author collaborations. Each node represents an author $\mathscr{A}_i$ and an edge between nodes $\mathscr{N}_{\mathscr{A}_i}$ and $\mathscr{N}_{\mathscr{A}_j}$ indicates that authors $\mathscr{A}_i$ and $\mathscr{A}_j$ co-authored a DFRWS EU paper. The colours represent communities with more than 7 members, per Blondel et al. (2008)'s community detection. The 10 largest are coloured, and the authors with the most publications (at least 5) are labelled (A–H). Node size is determined by the number of papers. (For interpretation of the references to color in this figure legend, the reader is referred to the Web version of this article.)

### 3.4. Best paper awards

Fig. 1 provides an overview of the distribution of the 16 Best Paper Awards[1] that were awarded since the establishment of DFRWS EU across our classification. While the data does not point to a singular research area being favoured by the forensic community, there is a noticeable concentration of awards in the fields of Digital Forensic Science, Device Forensics, and Memory Forensics, highlighting their significance in the community. Notably, areas such as Multimedia, or Artefact Forensics have not received awards.

## 4. Qualitative findings

The following subsections, corresponding to the green boxes in Fig. 1, organise the qualitative analysis into categories aiming to provide a thematic overview of DFRWS EU research.

### 4.1. Digital forensic science

This category encompasses general topics, including standardisation, legal and ethical considerations in digital investigations, and improving the reliability and resilience of digital investigations, education, and the investigative process. A total of 25 papers belong to this category – highlighting the importance of establishing foundational principles in this practitioner-oriented field.

Education-related articles mostly overlap with other Digital Forensic Science categories. Specifically, two articles presented data synthesis frameworks, EviPlant by Scanlon et al. (2017) and ForTrace by Göbel et al. (2022), which can be used to create scenarios (e.g., disk images, memory dumps). \Henseler and van Loenhout (2018) focused on education judges, prosecutors and lawyers (intersection of law & policy) while Hitchcock et al. (2016) discussed a training model for field triage for first responders. Lastly, Freiling and Zoubek (2017) conducted a student experiment among graduate students aimed at exploring factors, strategies, and tools that impact the outcome of an investigation.

Standardisation also received significant attention, although the contributions do not evenly spread over the years. Most of the work in this category targeted a unified representation of information. For instance, Cruz et al. (2015) presented a distributed data storage format that leverages AFF4 and can be used for incident response, Casey et al. (2015) explored CybOX™ for storing and sharing digital forensic information, and Stelly and Roussev (2018) presented Nugget a domain-specific language. Concerning setting standards, three articles were found. Amann and James (2015) aimed at building robustness and resilience investigation laboratories based on survey findings, while Nemetz et al. (2018) presented a corpus of SQLite databases, which may be used as a standardised dataset. Ottmann et al. (2022) outlined quality criteria for 'forensically sound' acquisition of main memory. Lastly, Park et al. (2018) conducted a case study on the data protection legislation and government standards to implement Digital Forensic Readiness as a mandatory requirement.

The digital forensic process as a whole was targeted by 12 publications. This category includes works targeting to improve the process or evidence handling. Thus, articles focusing on evidence validation, evaluation, interpretation, or attribution can be found here. Biedermann and Vuille (2016), Mutawa et al. (2016), Gruber et al. (2023a, 2023b) discussed various aspects of improving the handling, presentation, and understanding of digital evidence in forensic investigations. Topics included comprehensive evidence presentation, behavioural analysis, and addressing contamination risks. Schneider et al. (2021) complemented these thoughts by showcasing that a significant portion of low-cost USB flash drives sold as original contained non-trivial user data, highlighting the widespread issue of poor sanitisation practices and potential implications for forensic investigations. Karafili et al. (2020), Franqueira and Horsman (2020), Casey et al. (2020) introduced advanced tools and methods to aid forensic analysts in conducting investigations and producing high-quality reports. The research focused on argumentation-based reasoners and structured argumentation techniques to improve communication, quality, and analysis. Within the area of the digital forensic process is the work by van Baar et al. (2014) who described Digital Forensics as a Service (DFaaS) and thus a paradigm shift. Pringle and Burgess (2014) presented FCluster, which serves as middleware, offering a secure environment for forensic data processing, ensuring data integrity control rather than functioning as an application program.

Other papers focused on the uncertainty within digital forensic traces. For example, Sandvik and Årnes (2018) conducted an investigation into the reliability of clocks under low power conditions, and showed that this can cause clocks to adjust, but at present it is not possible to say how likely such behaviour is. Spichiger (2023) presented an approach to evaluate potentially unreliable location-based evidence gathered by mobile phones using a reference device. Lastly, one article fell into Law & Policy, namely the work by Sokol et al. (2020) who explored the use of IP addresses in court decisions, examining their role as digital evidence and issues related to anonymization.

### 4.2. Device forensics

The forensic analysis of digital devices has been present at each iteration since the first DFRWS EU. Mobile device forensics has focused on several platforms. Iqbal et al. (2014) addressed the acquisition and analysis of Windows tablet artefacts running the Windows RT operating system, a version of Windows 8 designed to run on ARM architectures. Minnaard (2014) analysed the memory of an Android phone, two wireless routers and two HP laptops, one running Windows and the

---

other Linux, to determine the longevity and types of traces left behind from radio signals of nearby and in range 802.11 networks in a range of environmental and configuration states. Saleem et al. (2016) looked at the application of multi-criteria decision analysis to ensure investigators chose the most appropriate tools. The work was evaluated by comparing two tools on two mobile devices in several different instances of illegal activity. Groß et al. (2021) proposed a method for the recovery of encryption keys on Android devices with file-based encryption. Jennings et al. (2023) examined the Apple Health database (from the Apple iPhone and Apple Watch), for inaccuracies and inconsistencies around time and location data in the context of workout activities.

Servida and Casey (2019) examined home-based IoT devices, e.g., alarms and surveillance cameras, and their associated smartphone applications to extend approaches for the IoT trace collection and analysis. While general mobile device forensics methods could be used, the authors found that IoT platforms sometimes require device-specific approaches. Torabi et al. (2020) used AI to detect compromised IoT devices, infer and monitor scanning and botnet campaigns, and discover DDoS victims based on backscattered packets towards the darknet.

Zhang et al. (2020) published a case study on reverse engineering the Mirai botnet command and control server-side executables and live processes that were extracted from the memory and disk images from a controlled installation. The authors summarised their findings in a road map for Mirai botnet server forensics to aid future investigations.

Safaei Pour et al. (2019) proposed a data-driven methodology, dubbed L1-PCA, to infer and characterise Internet-scale IoT exploitations. The authors correlated large volumes of network telescope data with IoT-specific information to identify patterns of IoT probing campaigns. The authors identified IoT-orchestrated campaigns using Shodan-searching for open resolvers that can be abused for performing amplification attacks.

Sandvik et al. (2022, 2023) published two papers on IoT forensics. The first paper introduced a model to quantify data volatility in IoT devices and implemented it on Contiki OS and the Coffee File System, approximating data lifetime based on system data writing processes. Building upon this, the authors extended their research to fog systems, highlighting challenges like identifying data processing nodes and the need for efficient triage methods. They proposed a technique for prioritising evidence collection in fog systems by calculating the weight of paths passing through nodes based on service probabilities.

Fukami et al. (2017) presented a study on NAND flash memory chip-off analysis that explored thermal-based chip-off methods' impact on raw bit error rates, showing an increase of up to 259 %. The paper introduced a mitigation strategy that leverages NAND manufacturers' read-retry mechanisms, reducing raw bit error rates by 94.6 %. van Zandwijk (2017) used bit-error statistics in NAND flash as a source of forensic information was evaluated. Files were added to USB drives through the controller at different retention times, examining how retention time differences were reflected in the statistics of bit errors.

Sayakkara et al. (2020) explored electromagnetic side-channel analysis for uninstrumented software activity monitoring on an IoT device. They utilised a HackRF software-defined radio to capture EM emissions from an Arduino Leonardo. Automated feature selection was employed to identify the most relevant input channels out of 20,000, reducing data for machine learning models by 99.5 % while maintaining over 93 % accuracy.

There have been three papers looking at the acquisition and analysis of evidence from entertainment devices. Boztas et al. (2015) investigated a Samsung smart television, Davies et al. (2015) looked at a Sony Playstation 4 running several firmware versions, and Barr-Smith et al. (2021) examined a Nintendo Switch. More recently, there have been two papers on automotive forensics. Gomez Buquerin et al. (2021) examined the feasibility of automotive forensics using a generic automotive forensic process on a state-of-the-art vehicle over an OBD diagnostics interface to collect data from ECUs and identified gaps that need to be addressed. Lee et al. (2023) offered approaches to data analysis and recovery of evidence from built-in video devices such as those used for dashcam footage in Kia, Hyundai, and Genesis vehicles. Bordjiba et al. (2018) presented a framework for the detection, mitigation, and investigation of phone call scamming campaigns based on complaints that originated at device end-points. Other niche areas of device forensics have included the extraction and analysis of data from a Programmable Logic Controller found in industrial control systems (Rais et al., 2022), and a forensic readiness framework for investigations following the sabotage of components produced during the 3D printing process (Rais et al., 2023).

### 4.3. Digital forensic techniques and fundamentals

Papers in this section represent techniques that are used in some phases of the investigative process: either improvements to acquisition, techniques used as part of the analysis or overall process improvements such as triage.

In terms of acquisition, Freiling et al. (2017) analysed the effect of abstraction layers on forensic evidence acquisition. The authors were the first to formally characterise the loss of digital evidence from only having access to data at higher levels of abstraction, e.g., cloud storage, as opposed to traditional low-level data acquisition, i.e., physical complete storage acquisitions – with the difference being referred to as the *semantic gap*. This semantic gap broadens further in virtualised environments where data is mapped from virtualised blocks to underlying storage blocks and in TRIM-enabled SSDs.

Alendal et al. (2018) published reverse engineering of Samsung's secure boot enforced common criteria mode and a corresponding technique to circumvent it to facilitate forensic acquisition from the device. Disabling this mode, featured on several Samsung devices, through the leveraging of a security vulnerability discovered as part of their research, the authors open the device up to direct storage and RAM acquisitions by misusing the devices' firmware update protocol.

Closser and Bou-Harb (2022) authored the first DFRWS paper to discuss the acquisition of forensic information from quantum mechanical computers. Their work introduced a live forensic recovery approach and a proof-of-concept experiment targeting quantum computer gates with the view to recovering binary values from a quantum system. The authors highlighted the unique forensic challenges posed by quantum computer forensics, including decoherence, entanglement and noise.

Zoubek and Sack (2017) proposed a selective deletion method for legal data protection in digital evidence acquisition and storage. This method, implemented as a plugin for the Digital Forensic Framework (DFF), allows the deletion of non-relevant or privileged data and associated metadata, preserving the authenticity and integrity of the evidence. The forensic examiner specifies the data to be removed. While the implementation focused on NTFS-structured devices, the authors suggested its extensibility to other file systems.

Several papers have discussed triage techniques to enhance the digital forensic process. Vidas et al. (2014) introduced OpenLV (formerly LiveView), a triage tool that maps a disk image to a virtual hard disk, bypassing login passwords for virtualisation and booting. This facilitates an easier review of the image content using the Windows OS to explore the data live, albeit virtualised. It is advantageous with specialist software or for creating suspect system screenshots. Hargreaves and Marshall (2019) presented a novel triage approach using synchronisation artefacts from one device to infer content or events on another, potentially inaccessible or absent device. Demonstrated with a prototype, SyncTriage, this technique aims to expedite investigations by partly bypassing the 'gain access' phase of the digital forensic process.

While SyncTriage attempts to bypass the problem of gaining access, other papers present techniques for breaking into encrypted data. Kanta et al. (2022) considered optimising word lists for performing dictionary attacks based on suspects' interests, and importantly also ranks the words in the wordlist to try to accelerate identifying the correct password. Bichara de Assumpção et al. (2023) provided a method for retrieving the BitLocker Volume Master Key from a TPM for data decryption.

In addition to triage, comprehensive investigation techniques have also been explored. Bollé and Casey (2018) advocated for searching links between artefacts in cases, demonstrated with email addresses and tested with 207 real cases. Open-source intelligence, often considered a related field, was discussed by Matthews et al. (2021). The study involved collecting, analysing, and presenting open data from Snapchat, with discussions on provenance and reliability. This prompts the question of integrating open-source intelligence within digital forensics, rather than it being a separate entity.

Other analysis techniques broadly relate to data recovery. Zoubek et al. (2016) discussed an approach for data recovery from RAIDs by automatically detecting the RAID parameters using block-level entropy measurement and generic heuristics. Both Dewald and Seufert (2017) and Bayne et al. (2018) considered data carving. The former uses a pattern-based data carving method to search for metadata structures of inodes in ext4 (discussed in file system artefacts in 4.4). The latter shows how GPUs can be leveraged to improve the performance of pattern matching operations and focuses on the carving application of pattern matching and provides a tool 'OpenForensics' to implement the concepts in the paper. Sester et al. (2021) investigated the related area of file type identification and surveyed methods available, which were divided into signature-based, statistical approaches, and computational intelligence approaches. It then goes on to advance this area with n-gram analysis based on support vector machines and neural networks. A technique was also presented by Wüst et al. (2017), which advances the ability to open recovered corrupted files by monitoring a viewer program's operation when opening non-corrupt files, and then instruments while opening one that is corrupt - referred to as a 'force open' approach, evaluated on PNG, JPG, and PDF files.

Nine papers have been published on hashing, approximate matching/fuzzy hashing, and perceptual hashing. Martín-Pérez et al. (2021) proposed a classification scheme for similarity digest algorithms, distinguishing them at three abstraction levels: bytewise, syntactic, and semantic. The former compares the byte sequences forming the digital artefacts, and the latter two utilise the artefacts' internal structures and contextual attributes, respectively. Martín-Pérez et al. (2021) also outlined potential attacks against these algorithms, including collision, inversion, and substitution attacks.

Approximate matching can be used in three contexts; searching, e.g., alternative/similar device detection, streaming, e.g., file transfer detection, and clustering, e.g., organising an unknown input set (Bjelland et al., 2014). Breitinger and Roussev (2014) proposed an automated approximate matching evaluation framework by relating approximate matching results to the longest common substring (LCS) using real-world data. LCS was proposed as the technique to be used for a ground truth determined by algorithmic means – facilitating the automated testing, evaluation and comparison of approximate matching algorithms.

Hashing and approximate matching are computationally intensive, with approximate matching being more demanding (Breitinger et al., 2014). Published work focused on improving the speed of these techniques in digital forensic investigations (Breitinger et al., 2014; Winter et al., 2014; Penrose et al., 2015; Liebler et al., 2019). Breitinger et al. (2014) introduced an approximate matching technique that reduces lookup complexity from $\mathscr{O}(x)$ to $\mathscr{O}(1)$ for known file matching. Winter et al. (2014) presented hash indexing strategies based on metric trees and locality-sensitive hashing. Penrose et al. (2015) outlined a fast triage contraband detection approach with 99.9 % accuracy, employing partial disk cluster sampling and pre-computed Bloom filters. Liebler et al. (2019) extended three approximate matching lookup strategies, i.e., hash-database for hash carving (hashdb), hierarchical Bloom filter trees (hbft), and (3) flat hash maps (fhmap), each with its advantages and disadvantages depending on the use case. Coates and Breitinger (2022) presented a novel approach for identifying document similarity using a fast estimation of the Levenshtein Distance based on compressing documents in signature representations and using those signatures for comparison. Liebler and Baier (2019) introduced apx-- bin, an approach combining cryptographic hashing and approximate matching for executable binary-related similarity. This addresses variations due to compiler settings and modifications, enhancing executable classification.

McKeown and Buchanan (2023) explored perceptual hashing for robust image signatures, considering features like histograms, statistical data, colour, texture, edge histograms, and frequency domain stats. They evaluated perceptual hashing methods, such as PDQ, Neuralhash, and pHash library, assessing their performance through Hamming distance scores between unrelated images and variants.

There are also various other techniques presented that broadly are categorised as 'behaviour analysis'. At a low level, Alrabaee et al. (2015) discussed a technique for detecting reused functions in binary code and provided a demonstration of identifying the RC4 function in *Zeus* reused from the *Citadel* malware. Other techniques include authorship verification, i.e., determining if two documents originate from the same author, and Halvani et al. (2016) presented a technique to group browser activity into 'sessions' which allows inference whether activity is one-off or repetitive (Gresty et al., 2016).

This shows that while some techniques receive significant attention, e.g., hashing, other practical areas have had fewer published papers at the conference, such as triage, gaining access to encrypted data, and behavioural analysis.

### 4.4. Artefact forensics

This section discusses papers that investigate the artefacts, or forensic traces left by some part of the technology stack (file systems, operating systems, applications). Despite it being a core area, only 14 papers were published.

Five papers were identified targeting file system artefacts. These papers cover both relatively new file systems but also include forensic advancements concerning well-established file systems. Lanterna and Barili (2017) discussed 'deduplicated' file systems and covered both *OpenDedup* and the *Windows 2012 Deduplication (W2012Dedup)*, including a deep dive into the hexadecimal of file chunk storage, and the higher-level implications for deleted file recovery. Prade et al. (2020) covered the less common file system ReFS and built on existing work to analyse the internal structures and develop a page carver for deleted data. The work includes an extension of The Sleuth Kit to include support for ReFS, providing practitioners with a practical means of accessing data within this file system (Dewald and Seufert, 2017). considered file carving on ext4 (also included in Sec 4.3). The authors used a pattern-based data carving method to search for metadata structures of inodes allowing improved recovery of data from ext4 file systems. Nordvik et al. (2019) focused on NTFS, specifically the Object ID Index, including multiple experiments with file creation, copying, moving etc. and shows how this artefact can be used to link external devices to computers to which it has been attached, detect manipulations, and show which boot session a file belongs to. Finally, a more general technique by Bahjat and Jones (2019), discusses a new method for assigning dates and times to file fragments based on the other data that they are surrounded by, further developing the digital stratigraphy technique.

A total of four papers on Operating System Artefacts were published, with the first of these appearing in 2019. Of the four papers, one had a focus on Windows forensics, two on macOS forensics, and one examined several Unix-based operating system artefacts including macOS. Atwal et al. (2019) conducted eight experiments on the macOS Spotlight to determine occasions when metadata records for deleted files persist in the metadata store, and when they are recoverable from the unused space on the filesystem. Palmbach and Breitinger (2020) focused on the Windows Operating System and its resilience to timestomping techniques. The scope of the experiments was limited to the use of three timestomping tools on the artefacts `$LogFile`, `prefetch`, `$USNJrnl`, `*.lnk` files and Windows event logs. From the experiments, the authors presented five rules for the detection of timestamp inconsistencies on Windows systems. A study of timestamp behaviour was conducted by Thierry and Müller (2022) on Linux, OpenBSD, FreeBSD and macOS to determine what timestamps are modified by a given operation. The authors also present a framework for testing POSIX compliance for a selection of software libraries and applications on the operating systems. Additionally, the authors examined the effects of timestomping on MACB timestamps on these systems. More recently, Joun et al. (2023) proposed a methodology to identify potential evidence spoliation by finding all available traces relating to deleted files on macOS systems, alongside a tool and released a dataset as part of the work.

Given the importance of understanding artefacts that are generated by the use of software, there are surprisingly few papers that provide an analysis of application-level artefacts. BitTorrent Sync is discussed by Farina et al. (2014). Schipper et al. (2021) covered the Riot.im application that uses the Matrix protocol. On the mobile side, van Zandwijk and Boztas (2019) examined Apple Health, focusing on determining the accuracy of steps and distances recorded, which involved the use of 600 trials by five subjects, including analysis of where on the body the device was located. Also in the mobile area, Horsman and Conniss (2015) take a more 'offence focused' approach considering driving offences, documenting artefacts such as CurrentPowerlog.powerlog on iOS to differentiate normally answered calls from those answered hands-free, or to determine if a device was unlocked, or that applications have been used.

Formalised and peer-reviewed artefact analysis is surprisingly scarce, likely due to the rapid evolution of artefacts and the difficulty in creating specific experimental data. However, some research does adopt a more general approach to artefacts, e.g., Choi et al. (2019) considered three messenger applications (KakaoTalk, NateOn, QQ messenger) but the focus is also on the encrypted backing store databases. This provides generalisable findings, but with those specific applications as examples, therefore providing both short and long-term contributions.

### 4.5. Multimedia forensics

Gloe et al. (2014) conducted a study on source device authentication in AVI and MP4-like video streams. They discovered unique characteristics across 19 digital cameras, 14 mobile phones, and 6 editing tools. These traits allowed them to attribute devices to specific makes and models based on variations in container formats, compression algorithms, acquisition parameters, and file structure.

Mullan et al. (2019) ran a two-year study on JPEG image header analysis. The authors focused on smartphone image provenance identification, covering various smartphone models, hardware iterations, firmware versions, and imaging applications. In 2020, they extended their work by creating an open-set model for camera-make prediction using JPEG header data (Mullan et al., 2020). The model achieved over 90 % accuracy with unmodified images and 55–75 % for post-processed images.

Two papers, DeepUAge (Anda et al., 2020) and Vec2UAge (Anda et al., 2021), contributed to the automation of Child Sexual Exploitation Material (CSEM) investigation through underage facial age estimation using deep learning techniques on the VisAGe dataset. DeepUAge achieved a mean absolute error (MAE) of 2.73 years in 2020, while Vec2UAge improved upon this in 2021, achieving an MAE of 2.36 years using facial embeddings and a bespoke neural network.

### 4.6. Memory forensics

Over the past decade, memory forensics research has been a fundamental and steady focus of DFRWS EU, accounting for approximately 14 % of all published papers in this domain.

Numerous studies addressed memory acquisition challenges. Among these, Bauer et al. (2016) investigated cold-boot attacks on DDR3 memory. They discovered that vendors' scrambling of memory to prevent undesired side effects posed challenges for forensic investigators, necessitating descrambling for detailed examination. In response, they proposed a technique requiring at most 128 bytes of plain text for content retrieval. A year earlier, Stüttgen et al. (2015) identified firmware rootkits as a significant threat – highlighting the potential of acquiring firmware through memory forensics, specifically by capturing specific firmware often overlooked by standard acquisition tools. A comprehensive evaluation of 12 memory acquisition methods spanning from user-mode tools to kernel-level, DMA, emulation, and virtualisation techniques was presented by Gruhn and Freiling, (2016). Using atomicity and integrity as benchmarks, their results highlighted the efficacy of user-mode tools for process memory, while cold-boot attacks and virtualization were best for full dumps. They also found that kernel-level methods often violated the integrity by altering the content within memory. While the software tools evaluated by Gruhn and Freiling (2016) only focused on Windows, Stüttgen and Cohen (2014) presented a Linux technique using a 'parasite' kernel module for version-independent memory acquisition – later added to the Rekall framework.

With respect to innovative acquisition methods, Latzo et al. (2020) introduced BMCLeech, a framework leveraging Baseboard Management Controllers (BMCs) – co-processors designed for server management. Although the method had constraints, like accessing only the initial 4 GiB of memory, it emerged as a notably stealthy memory acquisition technique. Zubair et al. (2022) explored the growing realm of Industrial Control Systems, presenting a novel method for acquiring memory from Programmable Logic Controllers, commonly found in this sector. Additionally, they detailed potential threats and their identification based on the retrieved memory. Recent research delved into memory acquisition techniques leveraging the UEFI (Tobias Latzo Florian Hantke and Freiling, 2021). The authors introduced a methodology and corresponding implementation, allowing investigators to mimic a cold-boot attack by executing the acquisition code straight from the UEFI, resulting in a notably high atomicity.

Accurate identification of operating system structures is essential for subsequent memory analysis. Cohen (2015) examined the variations of structures and offsets across different versions of Windows and its Kernel. After analysing multiple versions of the Kernel and a specific Windows driver, Cohen observed that while structure layouts remained largely consistent across versions, global constants showed significant variations. Within the Linux domain, Socała and Cohen (2016) introduced an automated profile generation methodology, which poses a challenge given the multitude of kernel versions and configurations to account for. Their innovative technique, integrated into the Rekall framework, harnesses the kernel's source code and real-time target system data to generate viable profiles.

Windows in-memory analysis research has been a focal point in recent years. Sylve et al. (2016) addressed the time-consuming issue of scanning memory for Windows kernel object allocations by limiting the scan to specific memory pages, identified through a special bitmap structure. However, this more performant method may miss artefacts outside the Kernel's current virtual address space. Otsuki et al. (2018) introduced a technique for reconstructing stack traces in Windows x64 memory dumps, with case studies on malware analysis. Fernández-Alvarez and Rodríguez (2023) discussed DLL injection in Windows, a malware capability, and proposed a solution to locate DLLs in a process's memory by combining pages of the same DLL across multiple processes and dumps. Uroz and Rodríguez (2020) explored the value and challenges of verifying digitally signed files extracted from memory in Windows memory analysis.

A novel technique was introduced for the stealthy extraction of TLS master secrets from memory (Taubmann et al., 2016). This method employed virtual machine introspection to capture memory snapshots, followed by a brute force strategy to locate the master secret within. A similar approach using virtual machine introspection for the extraction of SSH key material was presented by Sentanoe and Reiser (2022). However, their method requires knowledge about the memory layout of the data structures used by the SSH implementation. Diving into more specialised memory analysis, Pridgen et al. (2017) explored the recovery of artefacts from runtime environments, notably the HotSpot Java Virtual Machine. Additionally, Fernández-Álvarez and Rodríguez (2022) shed light on traces present in the memory of the Telegram Desktop Application for Windows, whereas Nissan et al. (2023) employed support vector machines to reconstruct query activities from memory snapshots. Awad et al. (2023) worked on the memory structure of a particular Programmable Logic Controller including forensic artefacts, establishing a research foundation for this progressively vital domain.

### 4.7. Network forensics

Research in cloud and network forensics, while not as dominant as other areas, has seen significant advancements. A central challenge is the amount of captured network traffic, often including irrelevant data to the investigation. Divakaran et al. (2017) addressed this issue by introducing a framework that identifies and correlates anomalous patterns indicative of malicious actions. Their approach does not require a learning phase and achieves high accuracy with a low false positive rate. Another strategy employing Behavioural Service Graphs to extract evidence from infected machines within a campaign is presented by Bou-Harb and Scanlon (2017). Moreover, Lamshöft et al. (2022) dived deeper into malware in network traffic, presenting a threat analysis centred around covert channels via syslog and port scans, while also introducing a deep convolutional neural network-based detection method. In addition to malware, Dinh et al. (2015) focused on spam. They presented a software framework that extracts features from emails, stores them in a central database and performs detection and categorisation into campaigns.

Turning to the intricacies of network forensics, Spiekermann et al. (2017) addressed the complications in capturing data within virtual environments, presenting a framework and tool for capture in OpenFlow-controlled networks.

Gugelmann et al. (2015) presented a tool for HTTP/S analysis. Harnessing correlation and frequent pattern detection, the authors first limit existing events, which are subsequently visualised to represent a timeline of HTTP and HTTPS activity.

The importance of cloud forensics was already highlighted by Roussev and McCulley (2016). The authors examined Google Docs artefacts, outlining the limitations of traditional forensic approaches in handling cloud-native artefacts. Additionally, Boucher and Le-Khac (2018) discussed the often-overlooked aspect of synced evidence and the ambiguity of artefact origin (local/synced) and proposed a novel framework as a first step and demonstrated its application to Google Chrome.

Recent research has ventured into the realm of cryptocurrency forensics. Thomas et al. (2022) introduced a blockchain query system designed from a forensic point of view, ensuring the forensic soundness of their implementation.

### 4.8. Miscellaneous

This section describes topics that could have been divided into other sections, but taken together, they represent an important topic area for discussion.

#### 4.8.1. Anti-forensics

Anti-forensics, i.e., discussing techniques and performing experiments that hamper investigations, has been identified as one of the miscellaneous subcategories with a total of six articles. Nilsson et al. (2014) discussed the possibility of hiding the full-disk encryption key, which is usually loaded into memory for ARM architectures to avoid cold boot attacks. Palutke and Freiling (2018) presented `Styx` which is a system hiding itself in memory and overcoming tools performing software-based memory acquisition. Göbel and Baier (2018) described hiding information, i.e., steganography, in the ext4 timestamp attribute. Freiling and Hösch (2018) (focus on disk) and Schneider et al. (2020) (focus on memory) conducted experiments where graduate students had to distinguish their forgeries from originals, and showed the difficulty of the manipulation task. In particular, it showed that manipulation is often more effortful than detection. Based on their findings, Schneider et al. (2022) focused on assessing the factors contributing to successful experiments in digital evidence tampering, drawing insights from previous studies and their experimental errors.

#### 4.8.2. Malware

The work in malware has largely focused on malware detection and code authorship. Alrabaee et al. (2014) presented a multi-layered approach to code authorship by examining software library functions, a code syntax dictionary and the way registers are manipulated. Based on this initial work, Alrabaee et al. (2019) produced BinChar for code authorship which used CNN and Bayesian probability. MalDozer, an Android tool for the detection and attribution of malicious applications using deep learning on API method calls, was presented by Karbab et al. (2018). 2019 also saw a paper on the MalDy system, a malware detection and threat attribution framework using supervised machine learning techniques (Karbab and Debbabi, 2019), a paper on the use of similarity hashing for classifying Window Portable Executable malware (Shiel and O'Shaughnessy, 2019), and work that provided a taxonomy of Windows OS Auto-Start Extensibility Points (ASEP) and a Volatility plugin called Winesap (Uroz and Rodríguez, 2019). An empirical study on the behaviour characteristic of the most prominent software supply chain attacks and an investigative framework that determines attack likelihood in a piece of distributed software was presented by Andreoli et al. (2023). Rathore et al. (2023) proposed Android malware detection models and a defence strategy against adversarial attacks known as MalV-Patch.

## 5. Future directions: Research and best practices

After a comprehensive analysis of ten years' worth of research and presentations at DFRWS EU, this section offers suggestions and recommendations for strategically advancing the digital forensics field in the evolving landscape of the future.

### 5.1. Artificial intelligence and digital forensics

AI's use in digital forensics is growing, though it is still very much in its early stages. Most existing research focuses on traditional AI applications – particularly clustering and classification tasks, but many digital forensic domains have not yet fully exploited AI's potential. There is a compelling need to explore diverse areas and integrate AI into various investigative stages (Du et al., 2020). The advent of Large Language Models, in particular, could significantly advance digital forensics (Scanlon et al., 2023). However, there is limited discussion on key digital forensic issues like explainability, which is crucial for AI's successful integration into digital investigations. Forensics of AI systems, as suggested by Baggili and Behzadan (2020) and Schneider and Breitinger (2023), is an intriguing, largely unexplored frontier within the DFRWS EU and broader digital forensic community.

### 5.2. Digital forensic datasets

Data and datasets are crucial in research, facilitating experimentation and ensuring result comparability and reproducibility (Garfinkel et al., 2009). A study by Grajeda et al. (2017) noted a positive trend towards researchers sharing their datasets, and these datasets being used by others – a finding echoed in this review.

However, it is often challenging to determine whether a dataset was reused or created, and if created, its availability. To improve discoverability, authors should clearly state these details in the abstract and keywords, using terms like 'dataset' or 'corpora'. This aids automatic data parsing, such as using the Elsevier API. Including datasets in NIST's CFReDS project at cfreds.nist.gov is also recommended, though the ability to directly cite individual datasets would be beneficial.

There's a need for mechanisms linking datasets with articles, enabling queries like "return all articles using dataset *X*". This is vital as each study provides insights into a comprehensive understanding. As Roussev (2011) emphasised, establishing ground truth for any non-trivial dataset is challenging, especially for an individual researcher.

### 5.3. Implementations/tools

Over the past decade, numerous techniques and frameworks have been introduced, often accompanied by implementations and tools for evaluation. Similar to the need for datasets, these must be accessible to the community. Wu et al. (2020)'s study found that only half of the tools developed in the research were publicly available, a pattern echoed in this study. The lack of a standardised protocol makes it challenging to ascertain if a tool has been published in a research paper. Echoing Wu et al. (2020), there is a clear need for a structured process for publishing tools. As suggested in Digital Forensic Datasets, developers should highlight their work in keywords/abstracts and release their tools as open source. A centralised repository for digital forensics tools could be one viable solution, where each tool is tested and documented before approval, which would benefit forensic practitioners and academia, strengthening future investigations and research alike.

### 5.4. Call for Papers, topics, and published work

The digital forensics research landscape constantly evolves, as reflected in the DFRWS EU Call for Papers (CfP). New topics of interest include cloud, covert channels (e.g., Tor, VPN), digital evidence sharing and exchange, digital forensic preparedness/readiness, implanted medical devices, SCADA/industrial control systems, smart power grid forensics, smart building forensics, vehicle forensics (e.g., drones, cars), and virtual currencies. Topics previously listed at a high level in 2014, such as application analysis, database forensics, digital evidence storage and preservation, filesystem forensics, multimedia analysis, traffic analysis, traceback and attribution, are now subcategorised under the areas above. Comparing the latest CfP with the presented topics at DFRWS EU, notably several CfP topics received minimal attention. This does not imply topic irrelevance, but rather presents research opportunities. Examples include smart power grid forensics, smart building forensics, and digital evidence and the law.

### 5.5. Inclusion and collaboration

Sec. 3 reveals diverse author origins and collaboration patterns, yet some European countries are unrepresented, and silos exist within the community. The under-representation of certain [European] countries indicates a need for broader international participation for a more inclusive global perspective. Despite growing international participation, many authors maintain isolated collaborations or consistent partnerships with few peers, as seen in the small grey clusters in Fig. 5. Research indicates that collaboration enhances productivity, as many studies require interdisciplinary, equipment-dependent, and project-based approaches (Lee and Bozeman, 2005). Therefore, increased collaboration and knowledge sharing can enrich the diversity of perspectives in the digital forensics community.

### 5.6. Research priorities and their balance

Fig. 1 reveals a discrepancy in authors' focus on different categories. For instance, Multimedia, and Network Forensics have fewer articles than Digital Forensic Science. This discrepancy might seem due to recurring authors dominating certain topics, but the decision on article acceptance rests with the technical program committee, not the authors. Thus, the research priorities and imbalances reflect broader trends in the digital forensics community. Furthermore, some categories, like multimedia forensics, have existing communities/venues. Another trend observed, reflected in the comparatively low number of articles in Artefact Forensics, is that the conference shifts to 'enduring work', i.e., work that remains relevant over time. For example, while an analysis of an application-specific artefact likely becomes invalid with the next application update, a novel standard or technique has longer durability. Other DFRWS initiatives such as DFIR Review[2] may also act as an alternative route for specific artefact research.

## 6. Limitations

This study's data collection and analysis largely relied on manual methods, with decisions influenced by the authors' expertise. While significant efforts were made to ensure consistency, it is acknowledged that different researchers might reach different conclusions based on their interpretations. Despite potential subjectivity, the authors believe most decisions were made carefully and reasonably.

The diversity of the articles posed a challenge, as many could not be clearly assigned to just one single category. The goal was to find the most fitting categorisation, but it is recognised that other researchers, including the article authors, may have categorised their work differently. This also applies to the classification in Fig. 1.

## 7. Conclusion

DFRWS EU has run for 10 years and has contributed significant research to the digital forensics community. By examining this single publishing venue, it is possible to gain insights into digital forensic research in Europe, and to some extent worldwide. Some areas are the focus of the research community and some areas form a critical part of the digital forensic process, e.g., file systems, and application artefacts that see less published research. In the Interpol review of digital evidence, Reedy (2023) states "digital forensics, now increasingly being referred to as digital forensic science, has reached a threshold of maturity both as computer science and forensic science". DFRWS EU consistently includes papers classified as 'digital forensic science', and these continue to formalise the discipline and improve the quality of results in the field. While this is essential for digital forensics to align with other forensic science fields, without this formalisation being complemented by peer-reviewed technical work including techniques that allow data to be extracted from data sources, and an understanding of artefacts that allow the interpretation of this data in the context of investigating crime, the technical capabilities within the field could formalise, but stagnate, risking missing important evidence as technology rapidly changes.

**CRediT authorship contribution statement**

All authors contributed in an equal manner.

**Declaration of competing interest**

The authors declare that they have no known competing financial interests or personal relationships that could have appeared to influence

---

the work reported in this paper.